\documentclass[aps,prl,twocolumn,superscriptaddress]{revtex4-1}
\usepackage{amsmath}	%allows the use of mathematical symbols
\usepackage{graphicx}	% allows the use of the include graphics functions
\usepackage{color}
\usepackage{gensymb}

\usepackage{hyperref}
\usepackage{upgreek}
\newcommand{\micro}{${\upmu}$}

\begin{document}

\title{Spin Textures of Polariton Condensates in a Tunable Microcavity with Strong Spin-Orbit Interaction }

\author{S. Dufferwiel}
\author{Feng Li}
\email[]{f.li@sheffield.ac.uk}
\author{E. Cancellieri}
\author{L. Giriunas}
\affiliation{Department of Physics and Astronomy, University of Sheffield, Sheffield S3 7RH, UK}
\author{A. A. P. Trichet}
\affiliation{Department of Materials, University of Oxford, Parks Road, Oxford OX1 3PH, UK}
\author{D. M. Whittaker}
\author{P. M. Walker}
\affiliation{Department of Physics and Astronomy, University of Sheffield, Sheffield S3 7RH, UK}
\author{F. Fras}
\affiliation{Department of Physics and Astronomy, University of Sheffield, Sheffield S3 7RH, UK}
\affiliation{IPCMS UMR 7504, CNRS and Universit\'e de Strasbourg, 67200 Strasbourg, France}
\author{E. Clarke}
\affiliation{EPSRC National Centre for III-V Technologies, University of Sheffield, Sheffield S1 3JD, UK}
\author{J. M. Smith}
\affiliation{Department of Materials, University of Oxford, Parks Road, Oxford OX1 3PH, UK}
\author{M. S. Skolnick}
\author{D. N. Krizhanovskii}
\email{d.krizhanovskii@sheffield.ac.uk}
\affiliation{Department of Physics and Astronomy, University of Sheffield, Sheffield S3 7RH, UK}

\date{\today}

\begin{abstract}
   We report an extended family of spin textures in coexisting modes of zero-dimensional polariton condensates spatially confined in tunable open microcavity structures. The coupling between photon spin and angular momentum, which is enhanced in the open cavity structures, leads to new eigenstates of the polariton condensates carrying quantised spin vortices. Depending on the strength and anisotropy of the cavity confinement potential and the strength of the spin-orbit coupling, which can be tuned via the excitonic/photonic fractions, the condensate emissions exhibit either spin-vortex-like patterns or linear polarization, in good agreement with theoretical modelling.

\end{abstract}

\pacs{}
%\keywords{}

\maketitle
Quantized vortices are topological defects occurring in many physical systems in optics, condensed matter, cosmology and fundamental particles, characterized by a phase winding of an integer multiple of $2\pi$ around a vortex core.  In semiconductor microcavities, quantized vortices \cite{Lagoudakis2008,Nardin2011,SanvittoD2011} and vortex-antivortex pairs \cite{Berloff2008,Cancellieri2014,Hivet2014,Christmann2012} may form spontaneously in exciton-polariton superfluids and non-equilibrium polariton Bose-Einstein condensates (BECs). Much effort has been devoted to the development of methods to create orbital angular momentum(vortices) in polariton condensates, providing ways to study the fundamental physics of metastable currents or for potential use as quantum sensors \cite{Franchetti2012} or information encoding devices \cite{Kapale2005}. Optical imprinting \cite{Krizhanovskii2010} of vortices as well as robust spontaneous vortices using chiral polaritonic lenses \cite{Dall2014} have been demonstrated. Interestingly, the coherent coupling of the photon pseudo-spin (polarization) with vortex orbital angular momentum has been shown to lead to new types of topological entities, named spin vortices, characterised by quantised spin current instead of phase winding. Uncontrolled spontaneous spin vortices were reported in atomic spinor BECs\cite{SpinorBEC} and in polariton condensates subject to structural disorder\cite{Manni2013}, although the exact origin of the polariton spin currents remains unclear. We also note that the degrees of freedom associated with both the orbital angular momentum and the polarization of a photon may find useful applications in quantum information processing  \cite{Mair2001,Nagali2009,Khoury2013}.

More recently, considerable attention has been focused on the investigation of polariton spin-orbit (SO) coupling, i.e. the interaction between the polariton orbital motion and its spin due to the effective magnetic field induced by the transverse-electric transverse-magnetic (TE-TM) splitting characteristic of semiconductor microcavities \cite{Panzarini1999}. In condensed matter SO coupling has led to significant physical phenomena  such as the spin-Hall effect \cite{Kato2004} and topologically protected conducting states \cite{Hasan2010}, whereas in optical microcavities, SO coupling of exciton-polaritons enables observations of interesting optical counterparts, including the optical spin-Hall effect \cite{Leyder2007}, magnetic-monopole-like half-solitons \cite{Hivet2012} and possibly topological insulators \cite{Lu2014,Nalitov2014}.

In this paper we demonstrate polariton condensation in our recently developed tunable open microcavity system \cite{Dufferwiel2014}, where a top concave mirror creates a zero-dimensional confinement potential for polaritons.  Multiple coexisting condensates are observed under non-resonant pumping exhibiting an extended family of spin vortices and textures. These effects are associated with the strong SO coupling in the open cavity system consisting of semiconductor bottom and dielectric top Bragg mirrors separated by an air gap. We observe condensate emissions showing both spin-vortex-like patterns as well as linearly-polarised states. The resultant condensate polarization patterns depend on the interplay between the strength and the anisotropy of the confinement potential and the strength of the SO coupling, which can be modified with change of exciton/photon fraction. We note that polariton condensates exhibiting less rich spin vortex phenomena were observed in a geometry of photonic micropillars coupled in a hexagonal pattern \cite{Sala2014}.

\begin{figure}
\center
\includegraphics[scale=0.95]{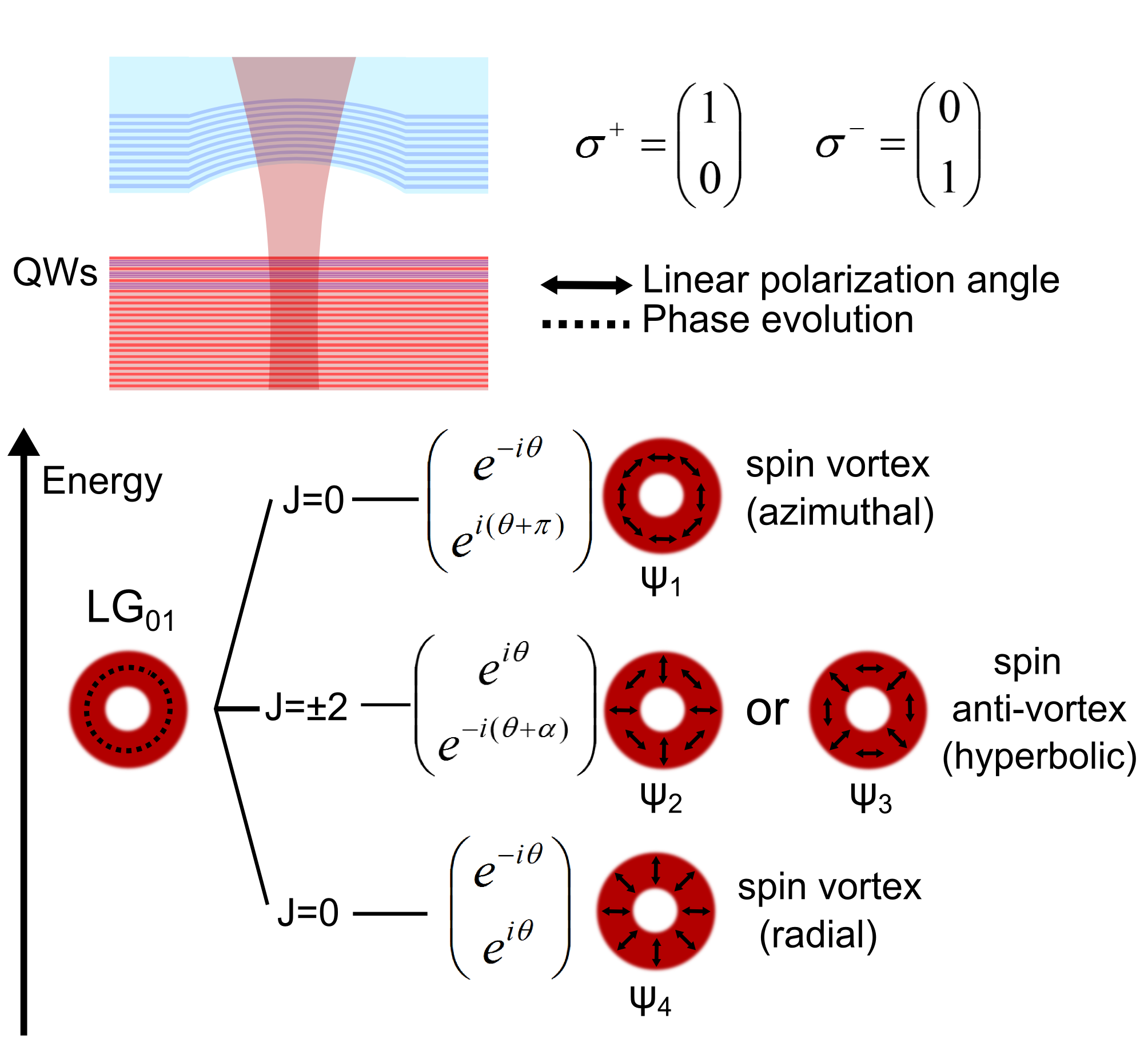}
\center
\caption{\label{fig1} Illustrative graph of the new eigenstates formed by $LG_{01}$ mode due to spin-orbit coupling. The TE-TM splitting in the cavity lifts the degeneracy of the modes leading to three energy levels characterised by spin vortices and anti-vortices. The mathematical form of the azimuthal part of each eigenstate is labeled next to its illustrative diagram. The top-left inset is a sketch of the open cavity. The basis of circular polarization is defined in the top-right inset.}
\end{figure}

The open microcavity system consists of planar bottom distributed Bragg reflectors (DBR) and a concave top DBR (see Supplementary Information \cite{Supplementary}) controlled independently by nanopositioners (top-left inset of Fig.~\ref{fig1}), which allows free tuning of the spectral resonance by changing the mirror separation \cite{Dufferwiel2014}. A total number of 12 GaAs quantum wells (QWs) are grown above the surface of the bottom DBR at electric field antinodes, allowing the strong exciton-cavity coupling regime to be reached with a Rabi splitting of $\sim15$ meV \cite{Supplementary}. Polariton condensation is demonstrated with nonlinear increase of emission intensity, sharp linewidth reduction and a small blueshift ($\sim1$ meV) far below the bare cavity mode at $\sim6$ meV to higher energy \cite{Supplementary}.

The top concave mirror induces a strong and almost harmonic lateral confinement of the polariton condensate\cite{Dufferwiel2014}. For this reason the system eigenmodes are studied in the basis of Laguerre-Gauss modes with the SO coupling or the asymmetries in the circular shape of the top mirror included perturbatively \cite{Supplementary}. In order to fully describe the eigenmodes of the system two bases of Laguerre-Gauss modes are needed, one for each pseudo-spin component: $LG_{pl}^{\sigma^{\pm}}$, where $\sigma^{+}/\sigma^{-}$ represent polaritons associated with left/right circularly polarized light, and $p$ and $l$ are quantum numbers quantifying the radial and azimuthal phase evolution, respectively.

We consider the simplest case of perfectly circular mirrors. Since only Laguerre-Gauss modes with $l\ne0$ carry non zero orbital angular momentum $\pm\hbar l$, corresponding to a phase rotation of $2\pi l$ either clockwise (phase vortex) or anti-clockwise (phase anti-vortex), our analysis starts from the first excited manifold (FEM) of the harmonic potential $LG_{0\pm 1}^{\sigma_{\pm}}$ (without SO coupling):
\begin{eqnarray}
LG^{\sigma^+}_{01}&=&C(r)\varphi_{11}(\theta)=\frac{r}{\sigma^2\sqrt{\pi}}e^{-r^{2}/2\sigma^{2}}\left( \begin{matrix}e^{i\theta} \\ 0 \end{matrix} \right)
\nonumber \\
LG^{\sigma^-}_{01}&=&C(r)\varphi_{-11}(\theta)=\frac{r}{\sigma^2\sqrt{\pi}}e^{-r^{2}/2\sigma^{2}}\left( \begin{matrix} 0\\ e^{i\theta} \end{matrix} \right)
\nonumber \\
LG^{\sigma^+}_{0-1}&=&C(r)\varphi_{1-1}(\theta)=\frac{r}{\sigma^2\sqrt{\pi}}e^{-r^{2}/2\sigma^{2}}\left( \begin{matrix}e^{-i\theta} \\ 0 \end{matrix} \right)
\nonumber \\
LG^{\sigma^-}_{0-1}&=&C(r)\varphi_{-1-1}(\theta)=\frac{r}{\sigma^2\sqrt{\pi}}e^{-r^{2}/2\sigma^{2}}\left( \begin{matrix}0\\ e^{-i\theta} \end{matrix} \right),
\nonumber
\end{eqnarray}

\noindent
where $\varphi_{sl}(\theta)$ is the azimuthal part of the polariton wavefunction with $s=\pm 1$ for polaritons associated with $\sigma^{\pm}$ polarized light, $C(r)$ is its radial part, and $\theta$ and $r$ are angular and radial coordinates. $\sigma=\sqrt{\hbar/m_{LP}\omega_{HO}}$, with $m_{LP}$ the lower-polariton mass and $\omega_{HO}$ the strength of the confining harmonic potential. Using degenerate perturbation theory and including the SO interaction, one obtains the following new eigenmodes (see \cite{Supplementary}):

\begin{align}
\psi_{1}(r,\theta) &=\frac{1}{\sqrt{2}} C(r)[\varphi_{1-1}(\theta)+\varphi_{-11}(\theta+\pi)]
\nonumber\\
\psi_{2}(r,\theta) &=\frac{1}{\sqrt{2}} C(r)[\varphi_{11}(\theta)+\varphi_{-1-1}(\theta]
\nonumber\\
\psi_{3}(r,\theta) &=\frac{1}{\sqrt{2}} C(r)[\varphi_{11}(\theta)+\varphi_{-1-1}(\theta-\pi)]
\nonumber\\
\psi_{4}(r,\theta) &=\frac{1}{\sqrt{2}} C(r)[\varphi_{1-1}(\theta)+\varphi_{-11}(\theta)]
\end{align}

\noindent
with eigenenergies: $E_{1}=E_{0}+\sigma^{2}\beta\pi$, $E_{2}=E_{3}=E_{0}$, and $E_{4}=E_{0}-\sigma^{2}\beta\pi$, where $E_{0}$ is the energy of the $LG_{01}$ mode and $\beta$ is a parameter describing the strength of the SO coupling \cite{Supplementary}. The structure of the new eigenmodes, illustrated in Fig.~\ref{fig1}, can be understood by observing that in the presence of SO coupling the conserved quantity of the system is the total angular momentum $J=l+s$. Since $l$ and $s$ are both equal to either $1$ or $-1$ the new possible eigenmodes have total angular momentum $J=-2, 0$ or $+2$. The SO coupling lifts the degeneracy by coherently combining the $J=0$ wavefunctions ($\varphi_{1-1}$ and $\varphi_{-11}$) to form new eigenstates, while leaving the energy of the two $J=\pm2$ modes unaffected. For these $J=\pm2$ any linear combination of $\varphi_{11}$ and $\varphi_{-1-1}$ is a suitable eigenmode in the presence of SO coupling.

\begin{figure}
\center
\includegraphics[scale=1]{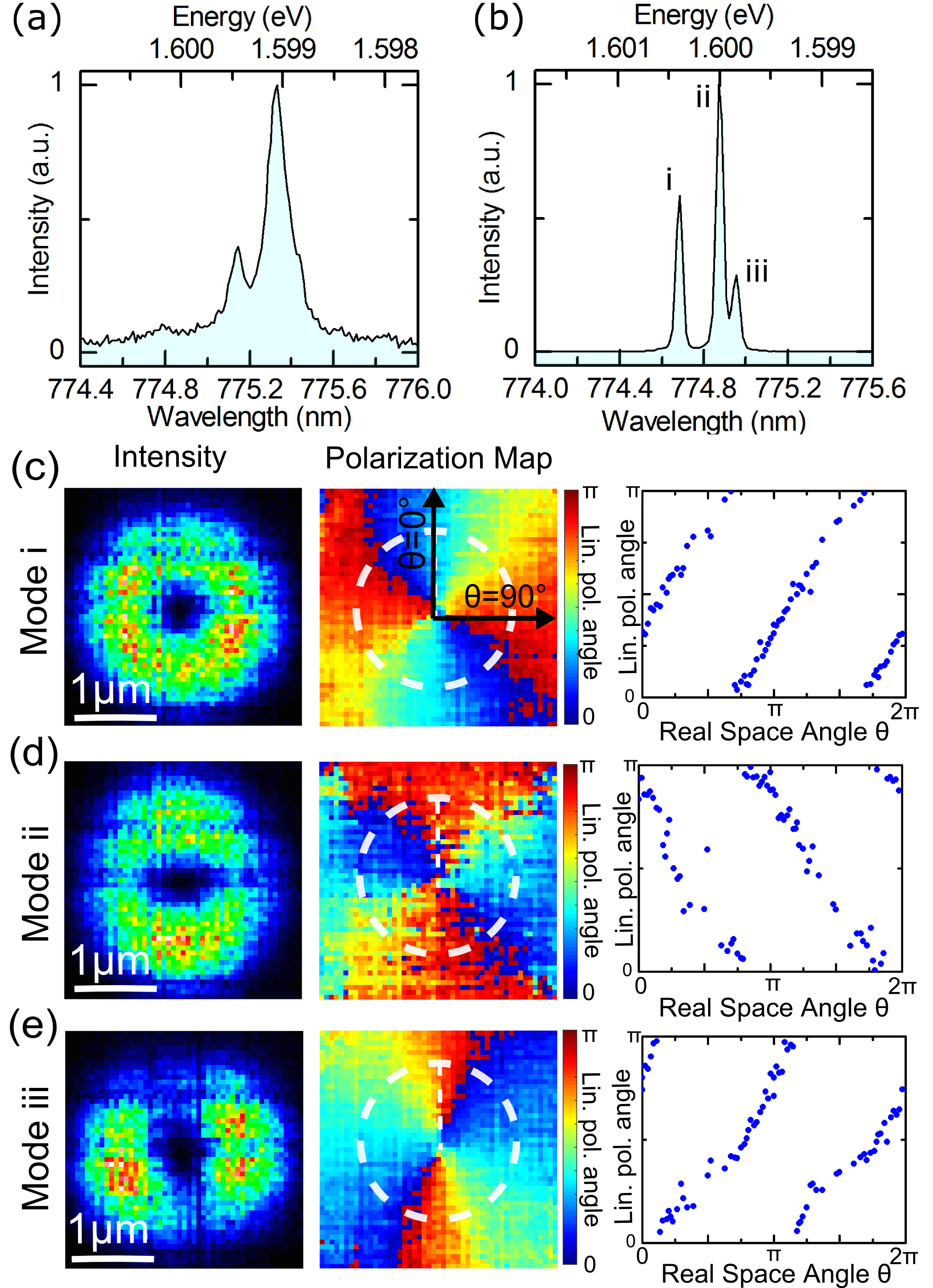}
\caption{\label{fig2} PL spectra of the FEM below threshold (a) and above threshold (b) where linewidth narrowing allows the resolution of three different modes. The RoC of the top concave mirror is 20 \micro m and the photonic fraction is 64\%. (c)-(e) Spatial intensity and polarization properties of mode i (c), ii (d) and iii (e). The left panels show the real space PL intensity; the middle panels show the real space distribution of the linear polarization angle; the right panels shows the linear polarization angle as a function of the real space winding angle circulating clockwise around the white dashed circle in the middle panel. $0^\circ$ and $90^\circ$ are defined in the middle panel of (c) which also defines the spatial coordinates used for all figures in this paper.}
\end{figure}

To investigate the properties of spin vortices low temperature photoluminescence (PL) measurements were carried out (details in \cite{Supplementary}). In the first set of measurements, a concave mirror with a radius of curvature (RoC) of 20 \micro m was employed and the mirror separation was $\sim1$ \micro m. The cavity is detuned so that polaritons in the FEM modes have a photonic fraction of $\sim64$\%. Below the condensation threshold, the spectrum associated with the FEM displays two broad features, as shown in Fig.~\ref{fig2} (a). With increase of pump power, condensation occurs and the linewidths drop sharply due to an increase of temporal coherence. Three well-resolved modes labelled by i, ii and iii are now revealed in Fig.~\ref{fig2} (b). Energy resolved images, shown in the left panels of Fig.~\ref{fig2} (c),(d) and (e), show a ring-like field distribution for all the three modes. The imperfection of the ring shape of mode iii is due to slight asymmetry of the confinement potential as will be discussed later. A linear polariser and a quarter wave plate are inserted into the optical path to collect polarization and energy resolved images for each mode in the horizontal-vertical ( $0^\circ/90^\circ$) basis, diagonal ($\pm45^\circ$) basis and circular ($\sigma^{+}/\sigma^{-}$) basis, and the associated Stokes parameters, $S_{1}$, $S_{2}$ and $S_{3}$, are calculated for each pixel of the image \cite{Supplementary}. The linear polarization angle $\phi$, defined as $2\phi=arctan(S_2/S_1)$, is mapped out for each mode in the middle panels of Fig.~\ref{fig2} (c),(d) and (e). As the circular polarization degree ($S_{3}$) is low for all the three modes \cite{Supplementary}, the linear polarization vectors characterize well the spin textures.

All three modes display quantised pseudospin currents characterized by a $2\pi$ rotation of $\phi$ around the mode cores, with a high linear polarization degree $\sqrt{S_{1}^2+S_{2}^2} \sim 0.95$ being exhibited. For both modes i and iii, $\phi$ changes nearly linearly with the real space azimuthal angle $\theta$, corresponding to the rotation of the vector of linear polarization clockwise around the mode centre, which indicates a co-rotating relation between $\theta$ and $\phi$, as indicated by the right panels of Fig.~\ref{fig2} (c) and (e). At $\theta=0\degree$, we observe $\phi=\sim90\degree$ (horizontal polarization) for mode i and $\phi=\sim0\degree$ (vertical polarization) for mode iii ($0\degree$ is defined as vertical, see the middle panel of Fig.~\ref{fig2} (c)), showing they are azimuthal and radial spin vortices corresponding to the extremal modes $\psi_{1}$ and $\psi_{4}$ in Fig.~\ref{fig1}, respectively. By contrast, mode ii is a spin anti-vortex displaying the opposite pseudospin vector rotation with respect to i and iii, with $\phi$ and $\theta$ counter-rotating (right panel Fig.~\ref{fig2} (d)). As discussed in \cite{Manni2013}, its hyperbolic-like polarization pattern results from the coherent combination, with any initial phase difference, of $J=\pm2$ half-vortices with different polarization (modes $\psi_{2}$ and $\psi_{3}$ in Fig.~\ref{fig1} correspond to the case with a phase difference of $0$ or $\pi$). The energy splitting of $\sim0.7$ meV observed between modes i and iii indicates strong SO interaction, consistent with transfer matrix simulations performed for the case of a planar open cavity revealing values of TE-TM splitting at high momenta of $\sim0.5$ meV. Such a large value mainly arises from the phase shifts due to reflections at the air gap interfaces in the open cavity system. Possible reasons for the unequal energy spacing between modes i, ii and iii are discussed in \cite{Supplementary}.

\begin{figure}
\center
\includegraphics{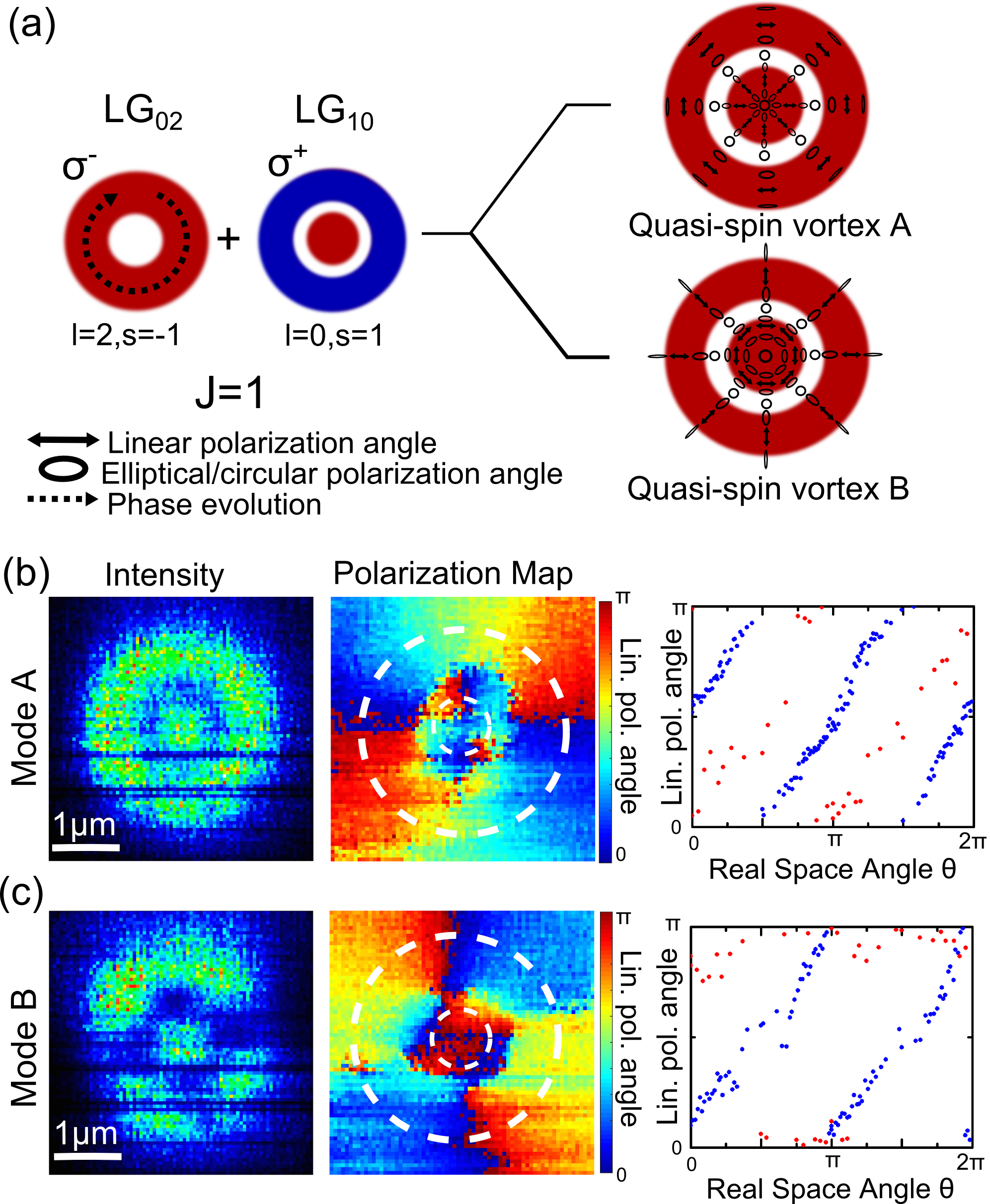}
\caption{\label{fig4} (a) Schematics of the generation of spin textures with higher order LG modes. (b) and (c) Experimental observation of the non-trivial spin modes in (a), with left, middle and right panels showing intensity profiles, linear polarization angle maps and polarization angle winding. The blue and red traces in the right hand panels show the outer and inner rotations defined by the dashed white circles in the middle panel.The horizontal dark lines on Figs 3(b) and (c) arise from dark pixels on the ccd detector.}
\end{figure}

We also observe spin textures for polaritons condensed into higher order LG-associated modes, when these are tuned into resonance with the exciton. Similar to the $LG_{0\pm 1}$ case, SO coupling also mixes modes in the second excited manifold like, for example, $LG_{10}$ and $LG_{02}$ modes.  As illustrated in Fig.~\ref{fig4} (a), the modes formed are quasi-spin vortices labelled as type A and B. The polarization vectors exhibit radial (A) or azimuthal (B) spin vortex character in the inner core and azimuthal (A) or radial (B) spin vortex character in the outer ring, connected by transient elliptically polarised states. Such quasi-spin vortices of polariton condensates were experimentally observed as shown in Fig.~\ref{fig4} (b) and (c), with a change of linear polarization angle of $\pi$ between the inner core and outer ring. Here above condensation threshold four spectrally resolved condensates are observed and for simplicity we show polarization patterns only for two of them, which fully demonstrate the principle illustrated in Fig.~\ref{fig4} (a). The imperfection of the mode spatial profile and the linear-like polarization vector of the inner core in Fig.~\ref{fig4} (c) compared to (a) are most likely due to the slightly elliptical shape of the top concave mirror.

\begin{figure}
\center
\includegraphics{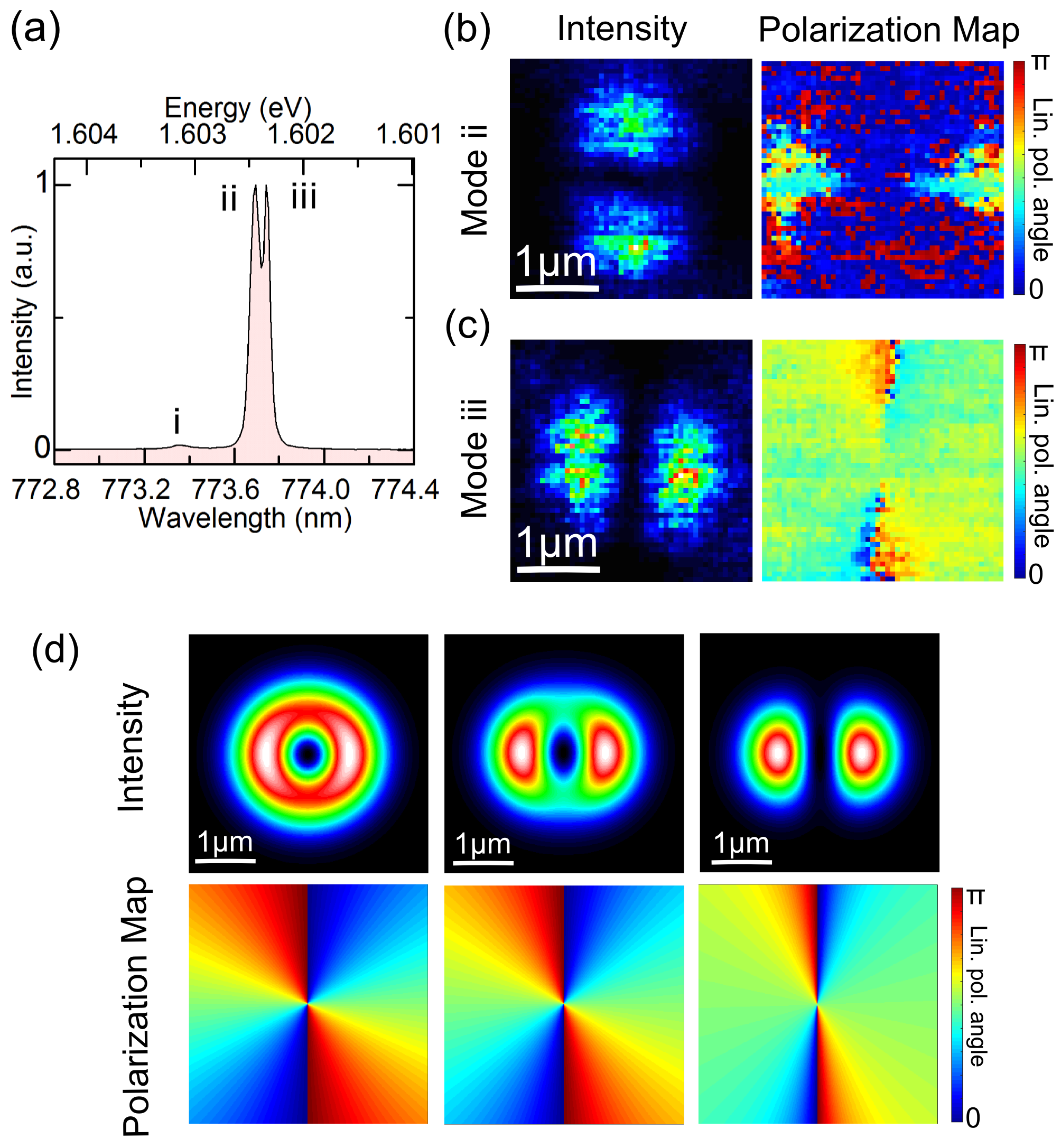}

\caption{\label{fig3} (a) Spectrum of the FEM above threshold with a photonic fraction of 41\% for a top concave mirror with RoC=$7$ \micro m. (b)(c) Real-space PL intensity profile (left panel) and real-space distribution of linear polarization angle derived from the Stokes parameters for modes ii and iii (right panel). (d) Simulation demonstrating the impact on the spin textures of the interplay between SO coupling and the cavity ellipticity. The ellipticity parameter $a$ is set to $-0.10$ $meV/\mu m^2$ for all graphs. Upper panels show real-space intensity profiles for SO coupling parameters $\beta=0.25, 0.05,$ and $0.01$ $meV\cdot\mu m^2$ from left to right, while the lower panels show the corresponding real-space maps of polarization angle. When the SO coupling decreases the mode changes gradually from a spin-vortex to a linear polarised MG-like profile. The definition of $\beta$ and $a$ is detailed in the supplementary materials \cite{Supplementary}. }
\end{figure}

If the concave top mirror has a sufficiently strong ellipticity which perturbs the harmonic confinement potential along the two orthogonal directions with strength $a$ (see \cite{Supplementary} for details) it may induce Mathieu-Gauss (MG) modes \cite{Vega2003, Nardin2010b} which are characterised by linearly-polarised orthogonal double-lobe profiles (See Fig.~\ref{fig3}(d)). The eigenmodes of the cavity arise from the competition between the asymmetry of the mirror and the strength of the SO coupling: either spin vortices or linear polarised states will be formed depending on which term dominates. In order to demonstrate condensation in MG modes the strength of SO coupling can be reduced by tuning the energy of the condensed modes closer to the exciton, and mirrors with smaller RoC chosen where the confinement potential is stronger and the spatial anisotropy is more pronounced.

From these considerations a concave mirror with a RoC of $7$ \micro m is chosen, and a photon fraction of 41\% employed. Fig.~\ref{fig3} (a) shows the spectrum of the polariton condensate associated with the FEM, where the low energy modes ii and iii are preferentially selected above threshold leading to significantly larger intensity than mode i. Non-trivial differences, compared to the spin vortices in Fig.~\ref{fig2}, are found in the mode spatial profiles and polarization patterns, as shown in Fig.~\ref{fig3} (b) and (c) for mode ii and iii. Instead of being spin vortices/anti-vortices, modes ii and iii clearly show MG-like orthogonal double-lobe profiles (left panels) with vertical (ii) and horizontal (iii) linear polarization (right panels). The simulated intensity distribution and polarization maps of one of the eigenstates confined in an elliptical potential are shown in Fig.~\ref{fig3} (f) for decreasing TE-TM splitting factors. Theoretically, it is seen that smaller SO coupling leads to the MG mode being the eigenstates of the system as the ellipticity term has greater influence. Importantly, for the same RoC=$7$ \micro m mirror, we can recover the vortex-like spin textures for all three modes similar to those shown in Fig.~\ref{fig2} by doubling the photonic fraction up to 82\%, as shown in the supplementary material \cite{Supplementary}. In addition, as expected this enhancement of SO coupling with increase of the photon fraction results in increase of the i-ii mode splitting from 0.71 meV to 1.02 meV. This demonstrates the advantages of the tunability of the open cavities in permitting the degree of the light/matter fractions of the cavity polaritons to be varied but also in allowing flexible manipulation of the condensate polarization textures.

In summary, we have demonstrated polariton condensate emission exhibiting spin vortices and more elaborate spin textures in a tunable microcavity system with lateral confinement. We note that below threshold the tomographic energy-resolved images of broad polariton modes also exhibit spin-vortex structures, but not all patterns are observed  due to the lack of spectral resolution. In order to reveal the complete mode structure it is important to be in the regime of condensation where the temporal coherence is increased. Our observations  are promising for optical information applications involving photon orbital angular momentum and polarization \cite{Kapale2005} and are interesting for further studies on cavity quantum electrodynamics (CQED) involving polaritonic spin.

We acknowledge support by EPSRC grant EP/J007544, ERC Advanced Grant EXCIPOL and the Leverhulme Trust.

\bibliography{manuscript}

%%%%%%%%%% Merge with supplemental materials %%%%%%%%%%

\widetext
\clearpage
\begin{center}
\textbf{\large Supplementary Information for ``Spin Textures of Polariton Condensates in a Tunable Microcavity with Strong Spin-Orbit Interaction''}
\end{center}
%%%%%%%%%% Merge with supplemental materials %%%%%%%%%%
%%%%%%%%%% Prefix a "S" to all equations, figures, tables and reset the counter %%%%%%%%%%
\setcounter{equation}{0}
\setcounter{figure}{0}
\setcounter{table}{0}
\setcounter{page}{1}
\makeatletter
\renewcommand{\theequation}{S\arabic{equation}}
\renewcommand{\thefigure}{S\arabic{figure}}
%\renewcommand{\bibnumfmt}[1]{[S#1]}
%\renewcommand{\citenumfont}[1]{S#1}
%%%%%%%%%% Prefix a "S" to all equations, figures, tables and reset the counter %%%%%%%%%%

\newpage
\section{Sample preparation and experimental set-up}
\begin{figure}
\center
\includegraphics[scale=1]{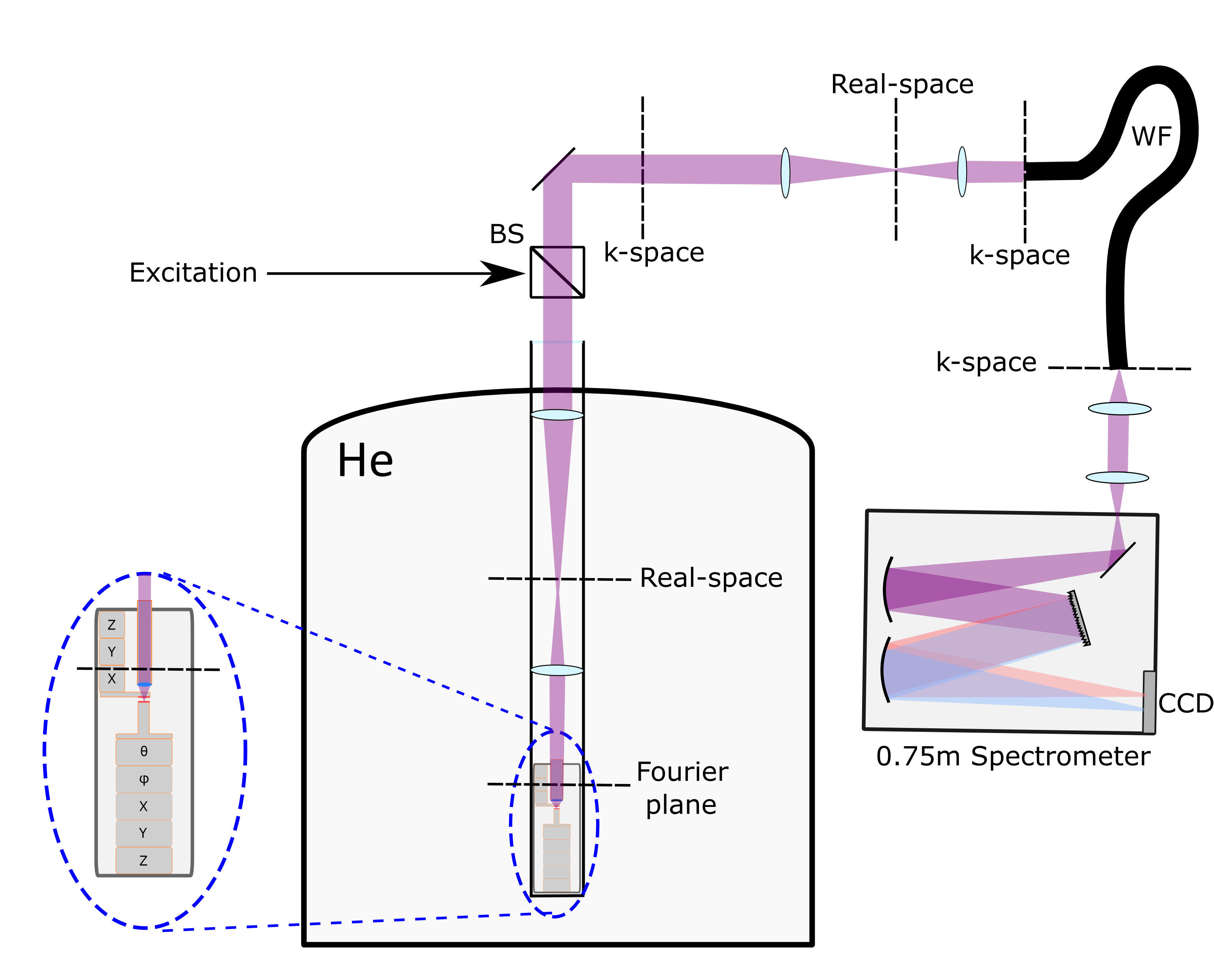}
\caption{\label{fig:setup} Schematic of the experimental setup.}
\end{figure}

The open microcavity system used to perform the experiments consists of a 31-pair Al$_{0.2}$Ga$_{0.8}$As/Al$_{0.95}$Ga$_{0.05}$As bottom distributed Bragg reflector (DBR) with a near-surface active region and an 11-pair SiO$_{2}$/TiO$_{2}$ circular-shaped concave top DBR separated by a micrometer sized gap. Arrays of concave mirrors are fabricated through focused ion beam (FIB) milling of a planar SiO$_{2}$ substrate before coating with dielectric layers \cite{Dolan2010}. The radii of curvature of the concave mirrors used in this work were 20 \micro m and 7 \micro m and can confine the polariton mode down to 1-2 \micro m. Nanopositioners allow independent positioning of both DBRs to form a planar-concave cavity where the spectral resonance can be tuned by changing the separation between them (Fig.~\ref{fig:setup}, left inset) \cite{Dufferwiel2014}. In the cavities used for this experiment, three sets of four 7 nm GaAs quantum wells (QWs) are embedded in the active region at the antinode of the optical field to enhance the exciton-photon coupling.

The experimental setup is shown in Fig.~\ref{fig:setup}. The cavity system is placed in a vacuum tube with a small amount of He exchange gas which is immersed in a liquid helium dewar. Optical access to the cavity is provided by placing an optical table on top of the dewar. The sample is non-resonantly excited at 630 nm, close to a stopband minimum, with a spot size of $\sim$ 30 \micro m on the top mirror surface. The beam is reflected into the dewar by a beam splitter (BS) to an objective lens above the sample (NA=0.55) and the photoluminescence from the cavity is collected along the same optical path. The final image is sent to the end facet of a wound fibre bundle (WF) consisting of a $4$x$4$ mm array of single mode fibres, which is imaged onto the spectrometer slits. It should be noted that light is depolarized by the WF, avoiding any measurement error possibly induced by the polarization-associated efficiency of the spectrometer mirror/gratings. Fig.~\ref{fig:setup} represents the setup used for k-space imaging, where lenses are located both in the dewar and on the optical table to form a confocal imaging system which projects the k-space image onto the WF facet. This setup can be easily altered for real-space imaging by simply replacing the two lenses on the optical table by one that focuses the real space image on to the WF facet, as used for our studies on spin vortices. A linear polarizer, whose polarization axis can be varied by motor-controlled rotation, is inserted before the WF to obtain polarization-resolved images, while adding a $\lambda/4$ plate enables acquisition in circular polarization basis.    

\newpage
\section{Demonstration of strong coupling}

\begin{figure}
\center
\includegraphics [scale=0.4] {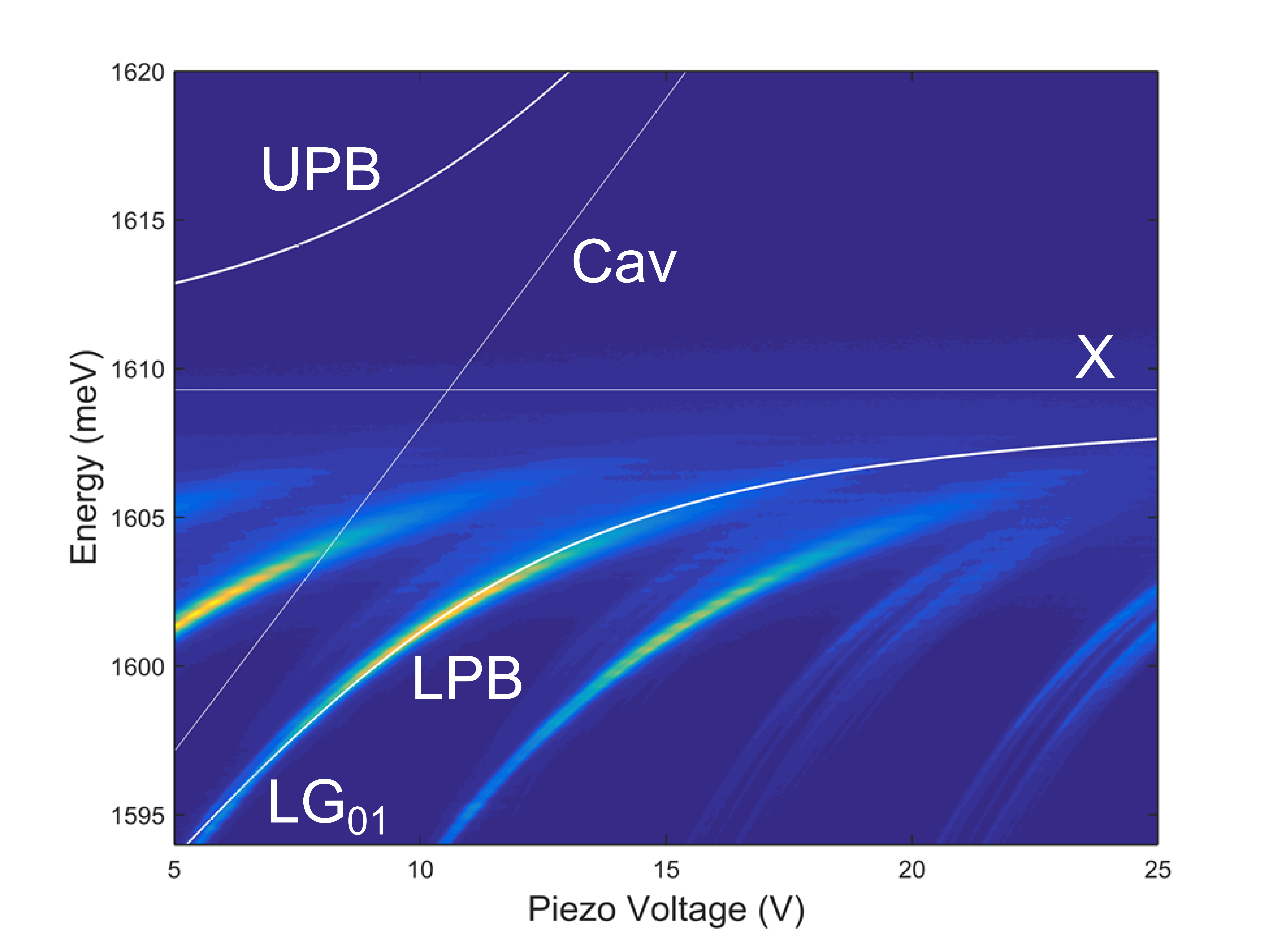}
\caption {\label{strongcoupling} Spectra of LPB for 0D cavity as a function of piezo voltage on bottom DBR. The simulated bare exciton (X), bare cavity (Cav), LPB and UPB dispersions are labelled, showing a Rabi splitting of 15 meV.}
\end{figure}

The strong coupling of the 0-dimensional (0D) cavity is demonstrated by scanning the PL spectrum with varying cavity length, as shown in Fig.~\ref{strongcoupling}. The cavity length is decreased by applying a DC voltage to the bottom z-nanopositioner, raising the bottom sample closer to the top mirror. The change in cavity length as a function of applied piezo voltage is closely linear for voltages less than $V = 20$ V. The lower polariton branch (LPB) shows strong curvature as a function of voltage arising from anti-crossing with the exciton energy, the signature of strong exciton-cavity coupling. The upper polariton branch (UPB) is not observed due to the strong absorption induced by the excitonic continuum of the 12 GaAs quantum wells. Fitting the LPB dispersion curve of the $LG_{01}$ mode with a coupled oscillator model gives a Rabi splitting of $\sim15$meV, in good agreement with theoretical and experimental values reported in similar structures \cite{Wertz2010,Bajoni2008}.

\begin{figure}
\center
\includegraphics[scale=0.3]{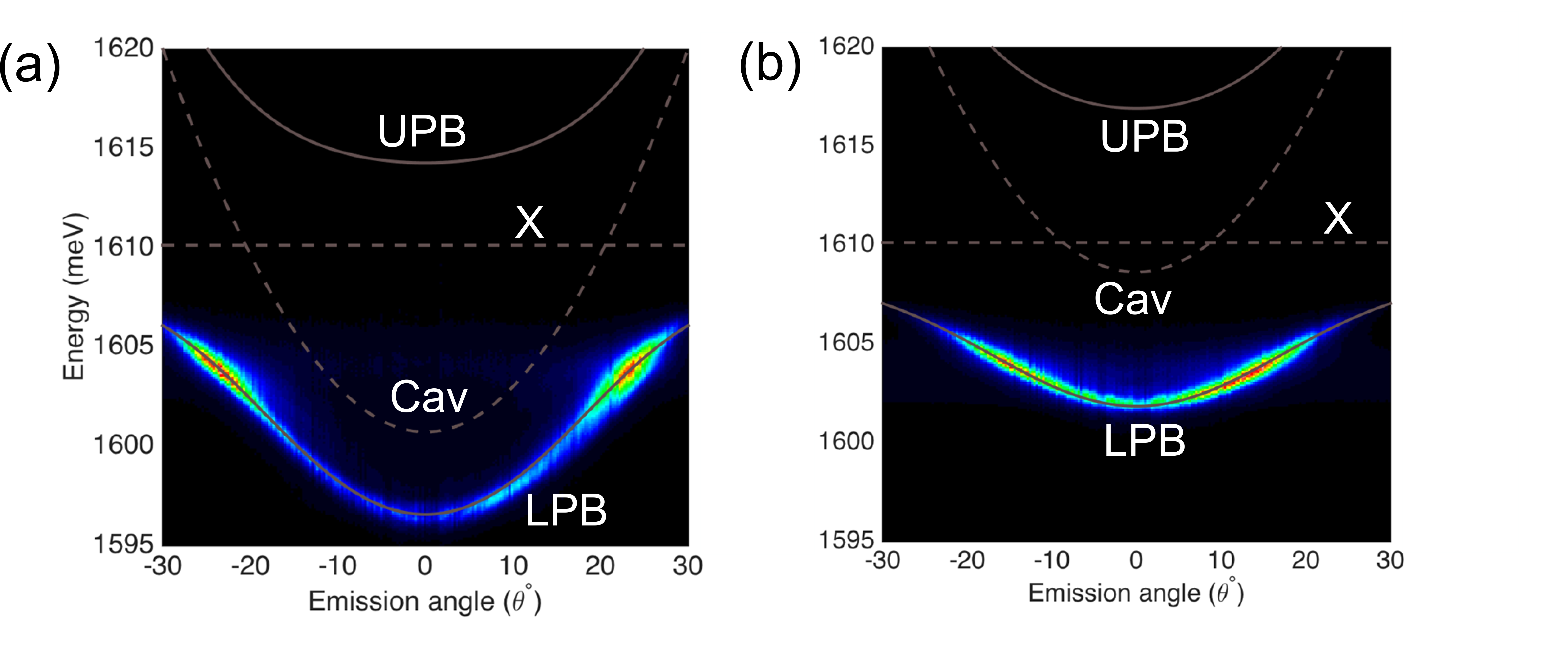}
\caption {\label{planar} Angular resolved spectra of LPB with planar top DBR for detunings of $- 9.4$ meV (a)  and  $-1.5$ meV (b). In both cases the simulated dispersions show a Rabi splitting of 15 meV.}
\end{figure}

The Rabi splitting can also be estimated from the 2-dimensional polariton dispersion. As shown in Fig. 1(a) of Reference~\cite{Dufferwiel2014} the top DBR contains planar regions around the arrays of concave features. The formation of a planar-planar cavity using this region gives rise to two-dimensional polaritons. Fig.~\ref{planar} (a) and (b) shows the angular dispersion of the planar open cavity at two-different detunings of $-9.4$ meV and $-1.5$ meV. In both cases the LPB can be fitted with the expected Rabi splitting of $15$ meV, in agreement with previous full microcavities containing similar numbers of QWs \cite{Wertz2010,Bajoni2008}. 

\newpage
\section{Polariton Condensation}
\begin{figure}
\center
\includegraphics{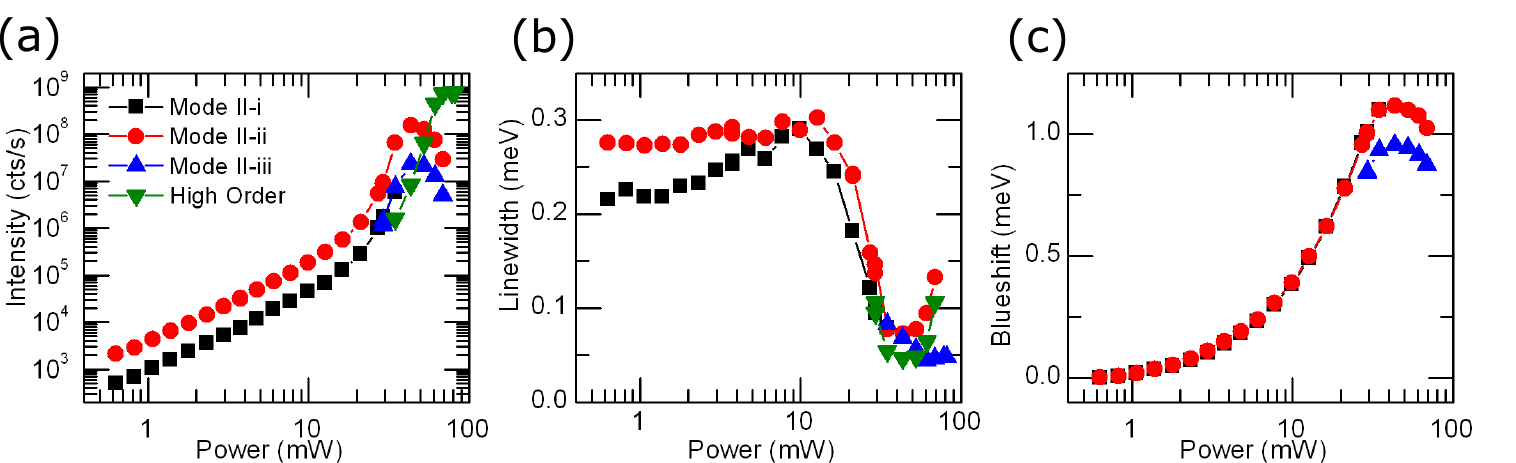}
\caption {\label{powerdep} Power dependence of the first excited manifold (FEM) of states at $\delta = -4 meV$, which splits into a triplet above threshold (black, red, blue traces), with spectrum in Fig.2 (b) of the main text. Integrated intensity (a), spectral linewidth (b) and energy blueshift (c) are shown for all the three modes. }
\end{figure}

Polariton condensation into the first excited manifold (FEM) of states is characterized in Fig.~\ref{powerdep}.  Non-linearity of output intensity and sharp linewidth reduction is observed. The blueshift at threshold, $\sim0.8$ meV, is significantly less than the $E_{cavity} - E_{LPB}$ which is $\sim6$ meV, showing that the cavity is in the strong coupling regime above threshold. The spectra below and above threshold are in Fig. 2 (a) and (b) of the main text.

\newpage
\section{Experimental determination of the Stokes parameters}

\begin{figure}
\center
\includegraphics[scale=0.7]{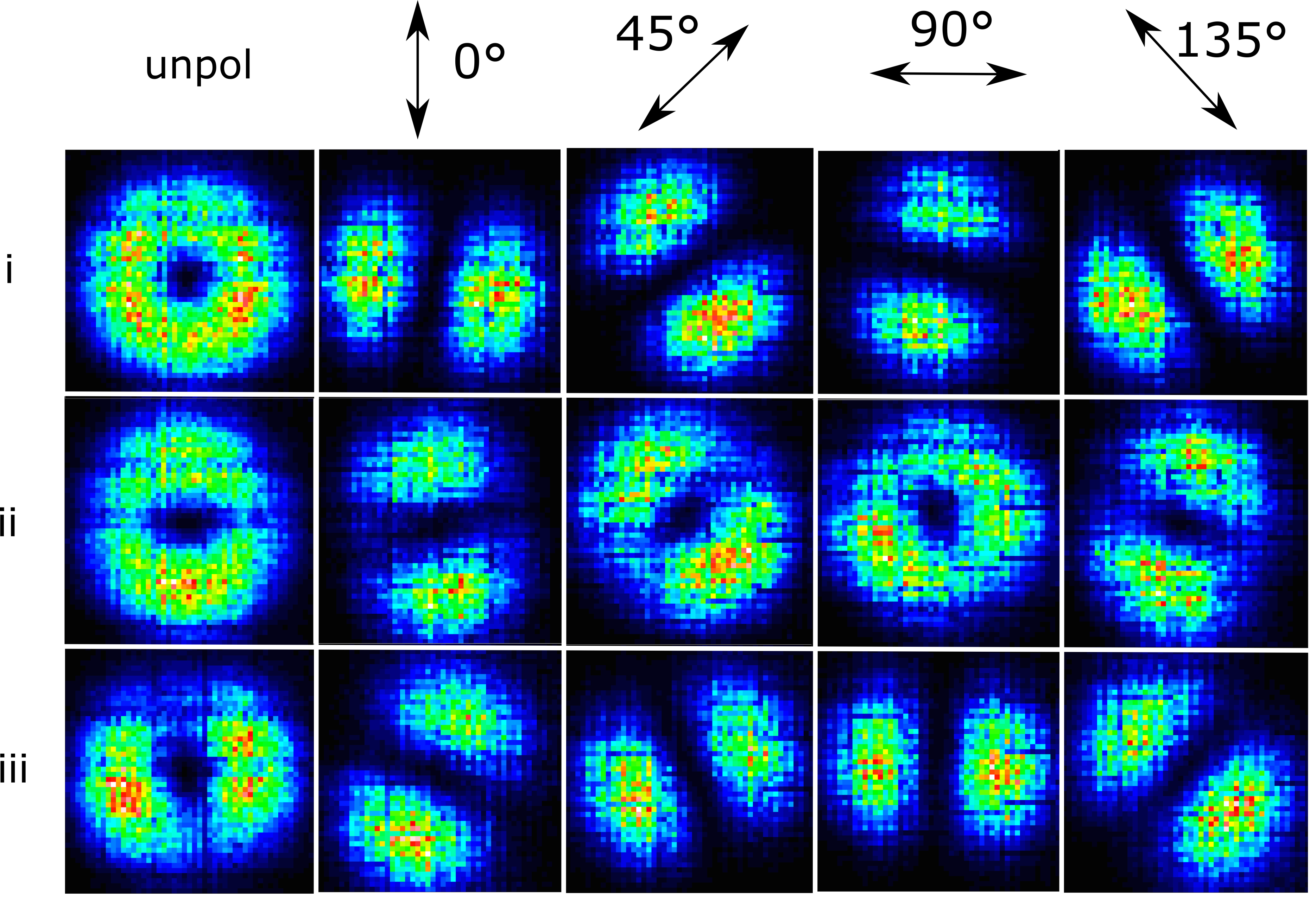}
\caption {\label{pol_resolved_image} Tomographic images of modes i, ii and iii in Fig. 2 of the main text, unpolarized (unpol), and with a linear polarizer with the polarization axis oriented at $0^\circ$, $45^\circ$, $90^\circ$ and $135^\circ$. }
\end{figure}

\begin{figure}
\center
\includegraphics {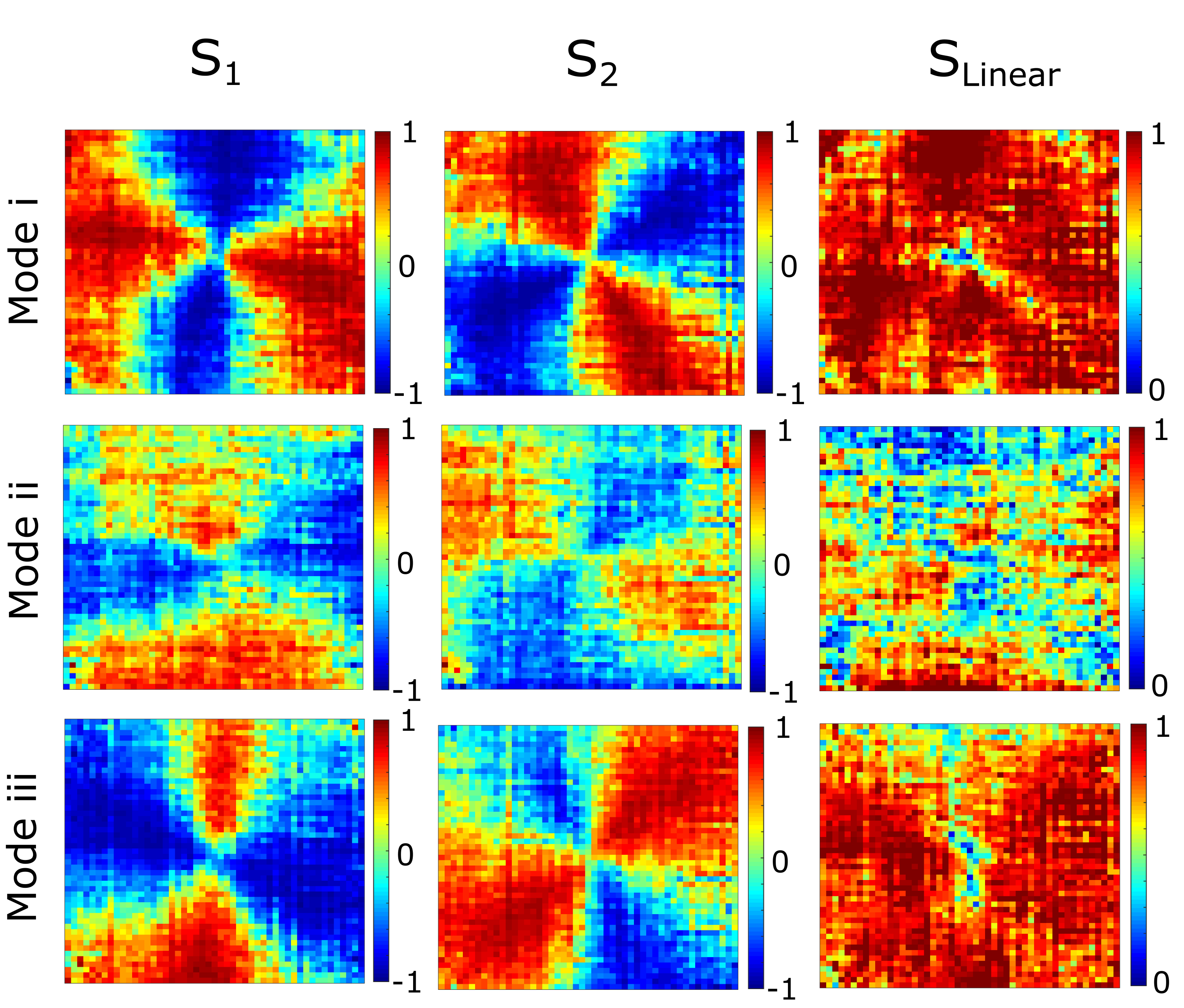}
\caption {\label{Stokes} Stokes parameters $S_{1}$ (a), $S_{2}$ (b) and $S_{Linear}=\sqrt{S_{1}^{2}+S_{2}^{2}}$ (c) derived for Modes i, ii and iii in Fig. 2 of the main text. } 
\end{figure}

In our experiment polarization-resolved images are taken in order to obtain the spatial intensity distribution in the horizontal-vertical ($0^\circ$ and $90^\circ$), diagonal ($45^\circ$ and $135^\circ$) and circular ($\sigma^+/\sigma^-$) bases. Fig.~\ref{pol_resolved_image} shows the clear rotation of the intensity distribution as a function of the linear polarizer angle. Modes i and iii show orthogonally positioned lobes co-rotating with the polarizer axis while mode ii anti-rotates with the polarizer axis. This indicates the formation of spin vortices for modes i and iii and spin anti-vortices for mode ii. Stokes parameters for horizontal-vertical ($S_{1}$), diagonal ($S_{2}$) and circular ($S_{3}$) basis are obtained for each spatial pixel of the mode by 

  \begin{align*} 
    S_{1} &= \frac{I(0^\circ)-I(90\circ)}{I(0^\circ)+I(90^\circ)} \\
    S_{2} &= \frac{I(45^\circ)-I(-45^\circ)}{I(45^\circ)+I(-45^\circ)} \\  
    S_{3} &= \frac{I(\sigma^{+})-I(\sigma^{-})}{I(\sigma^{+})+I(\sigma^{-})}
   \end{align*} 

where $I(\theta)$ is the measured intensity in various polarization basis. The polarization angle in real space $\phi$, is obtained from $2\phi=arctan(S_{2}/S_{1})$, as the winding angle in the $S_{1}$-$S_{2}$ plane of the Stokes presentation is twice of that in real space. Fig.~\ref{Stokes} shows the Stokes parameters derived for the spin vortices in Fig. 2 of the main text. 

\newpage
\section{Spin vortices with a top mirror of ROC=$7$ \micro m}
\begin{figure}
\center
\includegraphics [scale=1.2] {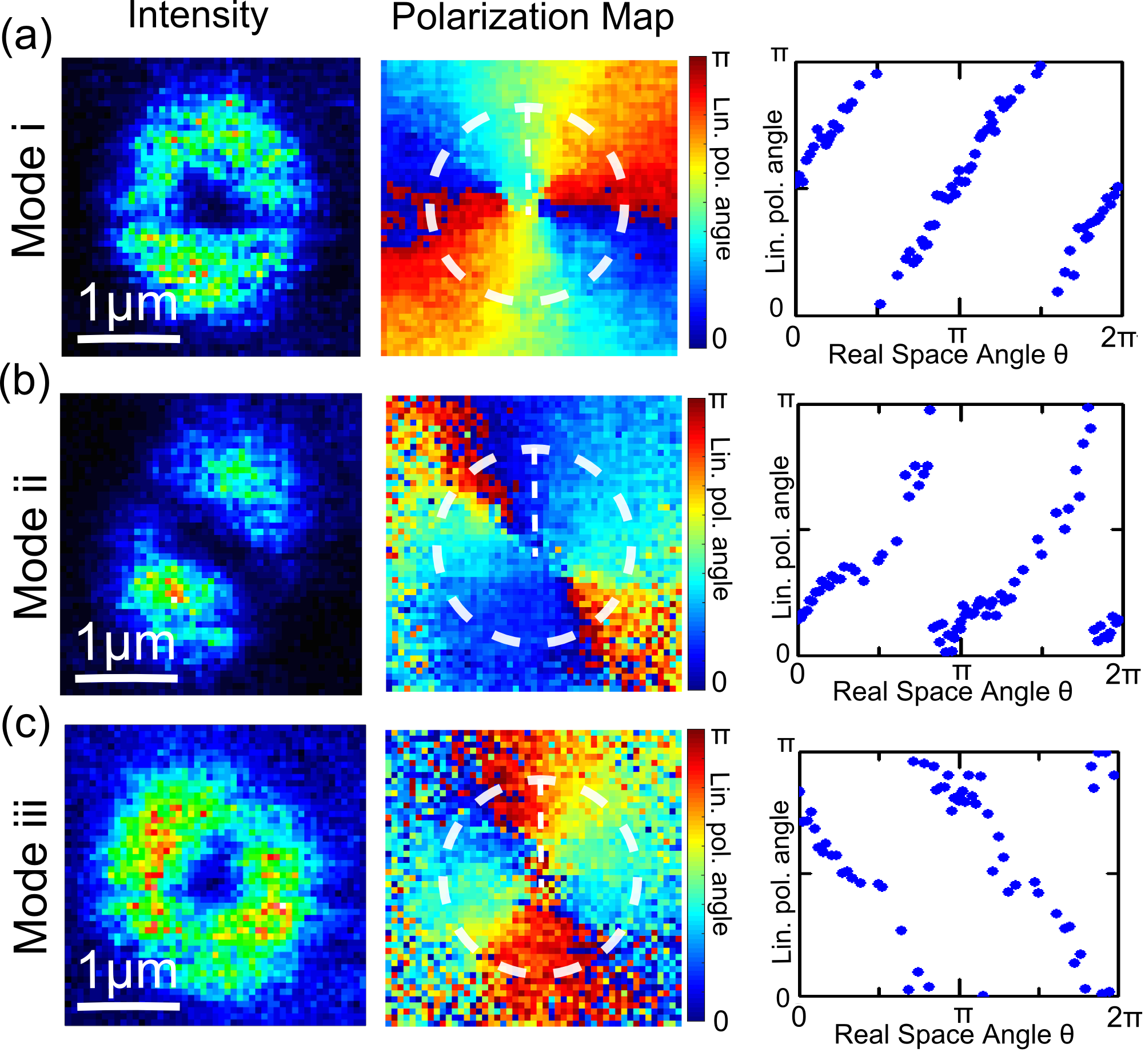}
\caption {\label{ROC7} Spatial intensity and polarization properties of the three eigen modes with a RoC=$7$ \micro m top mirror, shown in (a),(b) and (c) respectively for a photonic fraction of ~82\%. The left panels show the real space PL intensity; the middle panels show the real space distribution of the linear polarization angle; the right panels show the linear polarization angle as a function of the real space winding angle circulating clockwise around the white dashed circle in the middle panel. } 
\end{figure}

The mode spatial profiles and polarization patterns with the RoC=$7$ \micro m top mirror with a high photonic fraction (82\% photon-like) are shown in Fig.~\ref{ROC7}. Azimuthal, radial and hyperbolic spin vortices are revealed, similar to Fig. 2 but contrasting sharply with Fig. 4 of the main text taken at 41\% photon fraction where the ellipticity perturbation plays a larger role relative to the SO interaction. 

\newpage
\section{Theoretical Model}

We present here the theoretical approach used to interpret the experimental data. In our approach we use degenerate perturbation theory to find the eigenmodes of the polariton system in the case of low polariton densities (i.e. when the nonlinearities play a negligible role) in the presence of SO coupling, elliptical shape of the top concave mirror and birefringence from the anisotropy of the refractive index of the top mirror. Taking the eigenvectors $\left(\begin{matrix} 1 \\ 0 \end{matrix}\right)$ and  $\left(\begin{matrix} 0 \\ 1 \end{matrix}\right)$ to represent $\sigma^{+}/\sigma^{-}$ circularly polarised polaritons, the 2x2 Hamiltonian describing the lower-polariton branch in the linear regime can be written as:

\begin{equation}
H=
\left(\begin{matrix}
-\frac{\hbar^2\nabla^2}{2m_{LP}} + V & \beta\left(\frac{\partial}{\partial x} - i\frac{\partial}{\partial y}\right)^2+\Omega e^{i\theta}/2
\\
\beta\left(\frac{\partial}{\partial x} + i\frac{\partial}{\partial y}\right)^2+\Omega e^{-i\theta}/2 & -\frac{\hbar^2\nabla^2}{2m_{LP}} + V 
\end{matrix}
\right),
\label{eq1}
\end{equation}

\noindent
where $m_{LP}$ is the lower-polariton effective mass. The terms depending on $\beta=\hbar^2(1/m^{TE}_{LP}-1/m^{TM}_{LP})/2$, where $m_{LP}^{TE/TM}$ are the lower-polariton masses in the TE/TM polarizations, describe the TE-TM splitting. As pointed out in \cite{Dufferwiel2014} the top concave mirror induces a strong near-harmonic lateral confinement potential. $V=\frac{1}{2}m_{LP}\omega^2_{HO}[x^2(1+\delta)+y^2(1-\delta)]$, where $\omega_{HO}$ is the strength of the harmonic confinement. The terms $\pm a=\pm\frac{1}{2}m_{LP}\omega^2_{HO}\delta$ account for an elliptical asymmetry of the top circular mirror with the long axis either aligned along the $x$ or $y$ directions. As birefringence may arise in both the top and bottom mirrors due to strain, the terms $\Omega e^{\pm i\theta}/2$ account for a birefringence that induces a shift at $k=0$ between the TE-TM branches and tends to align the field polarization along the direction $\theta$. 

Since the harmonic confinement is much stronger than the SO coupling, the birefringence and the asymmetry, one can treat these terms as perturbations. To study a 2-dimensional harmonic oscillator several equivalent eigenvector bases can be used. Among them two are particularly useful: the basis of Laguerre-Gauss modes $LG_{pl}^{\sigma^{\pm}}$ (where $p$ and $l$ are radial and azimuthal quantum numbers) and the basis of Hermite-Gauss modes $HG_{sr}^{\sigma^{\pm}}$ (where $r$ and $s$ are quantum numbers along the $x$ and $y$ axes). While the basis of LG modes allows a more intuitive understanding of the shape of the spin vortices, the basis of the HG modes allows an easier evaluation of the matrix elements needed to determine the perturbed eigenenergies and eigenmodes. For this reason, and since the perturbed eigenmodes and eigenenergies do not depend on the basis of the Hilbert space used to evaluate them, we use the basis of the HG modes to apply perturbation theory. In the case of the first excited manifold the four relevant HG modes are:

\begin{eqnarray}
\psi_{1}(x,y)=\frac{x e^{-\frac{x^2+y^2}{2\sigma^2}}}{\sqrt{\sigma^4\pi/2}} 
\left(\begin{matrix} 1\\ 0\end{matrix} \right)
&  \hspace{1cm}        &
\psi_{2}(x,y)=\frac{x e^{-\frac{x^2+y^2}{2\sigma^2}}}{\sqrt{\sigma^4\pi/2}}
\left( \begin{matrix} 0\\ 1\end{matrix} \right)
\nonumber
\\
\psi_{3}(x,y)=\frac{y e^{-\frac{x^2+y^2}{2\sigma^2}}}{\sqrt{\sigma^4\pi/2}}
\left( \begin{matrix} 1\\ 0\end{matrix} \right)
&          &
\psi_{4}(x,y)=\frac{y e^{-\frac{x^2+y^2}{2\sigma^2}}}{\sqrt{\sigma^4\pi/2}}
\left( \begin{matrix} 0\\ 1\end{matrix} \right).
\nonumber
\end{eqnarray}

\begin{figure} [h!]
\center
\includegraphics[scale=1.0]{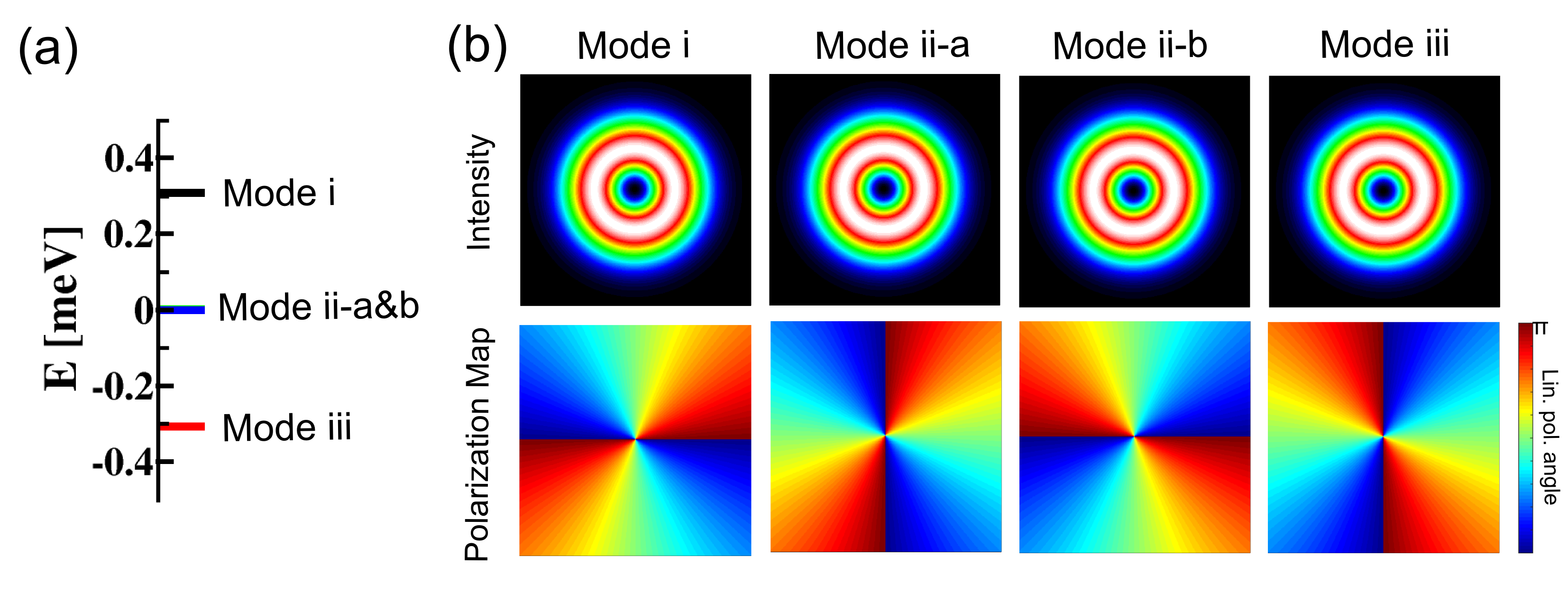}
\caption {\label{symmetric} Case for no mirror ellipticity and birefringence equal to zero. (a) Simulated energy levels: mode i (black), mode iia (blue), mode iib (green), and mode iii (red), the energy is evaluated with respect to the unperturbed mode. (b) Simulated polariton density (first row) and angle of linear polarization $arctan(S_{2}/S_{1})$ (second row). To obtain both the spectra and the eigenmodes the following parameters are used: $\beta=0.2$ $meV\cdot\mu m^2$, $\sigma=0.7$ $\mu m$. All graphs are $4\mu m \times 4\mu m$ in size.}   
\end{figure}

Using these modes as basis, the new perturbed eigenenergies and eigenmodes of the system are obtained by diagonalising the following matrix:

\begin{equation}
M=\left(
\begin{matrix}
\frac{1}{2}a\pi\sigma^6 & \frac{1}{4}\pi\sigma^2(-2\beta+e^{i\theta}\sigma^2\Omega) & 0 & \frac{i}{2}\pi\beta\sigma^2 \\
\frac{1}{4}\pi\sigma^2(-2\beta+e^{-i\theta}\sigma^2\Omega) & \frac{1}{2}a\pi\sigma^6 & -\frac{i}{2}\pi\beta\sigma^2 & 0 \\
0 & \frac{i}{2}\pi\beta\sigma^2 & -\frac{1}{2}a\pi\sigma^6 & \frac{1}{4}\pi\sigma^2(2\beta+e^{i\theta}\sigma^2\Omega) \\
-\frac{i}{2}\pi\beta\sigma^2 & 0 & \frac{1}{4}\pi\sigma^2(2\beta+e^{-i\theta}\sigma^2\Omega) & -\frac{1}{2}a\pi\sigma^6 \\
\end{matrix}
\right)
\end{equation}

\noindent
where $\sigma=\sqrt{\hbar/m_{LP}\omega_{HO}}$. The eigenmodes and eigenenergies of this matrix reduce to those of Equation (1) in the manuscript in the case of zero birefringence and no asymmetry in the harmonic confinement. For this particular case the energy spectra, the polariton density and the polarization angle for the four eigenmodes are plotted in figure ~\ref{symmetric}. As expected, the higher (black) and the lower (red) modes are azimuthal and radial spin-vortices respectively, while the two remaining central modes (green and blue) are spin-antivortices in agreement with the experimental observations in fig 2 (c,d,e) in the main text.

\begin{figure} [h!]
\center
\includegraphics[scale=1.0]{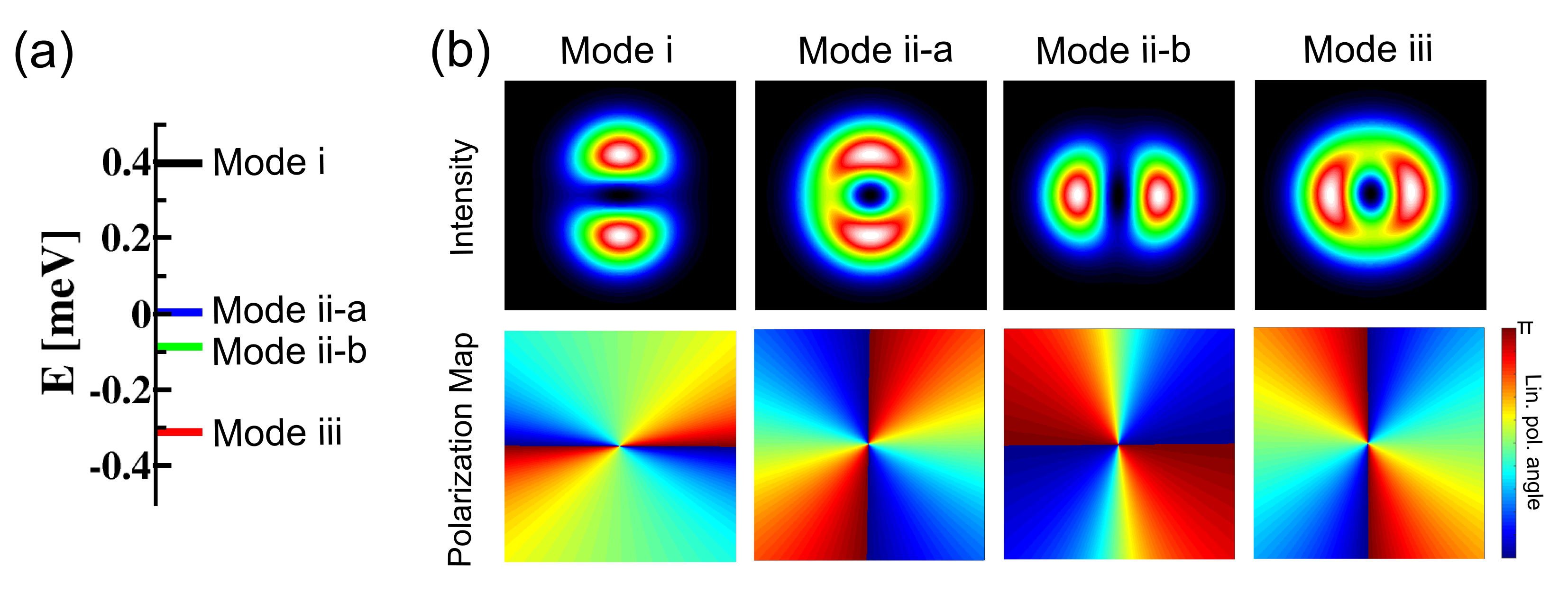}
\caption {\label{asymmetric} Case with mirror ellipticity and birefringence different from to zero. (a) Simulated energy levels: mode i (black), mode iia (blue), mode iib (green), and mode iii (red), the energy is evaluated with respect to the unperturbed mode. (b) Simulated polariton density (first row) and angle of linear polarization $arctan(S_{2}/S_{1})$ (second row). The following parameters are used: $\beta=0.2$ $meV\cdot\mu m^2$, $\sigma=0.7$ $\mu m$, $\Omega=0.4$ $meV$, $\theta=0.01\pi$, and $a=-0.6$ $meV$to obtain both the spectra and the eigenmodes. All graphs are $4\mu m \times 4\mu m$ in size.}   
\end{figure}

The case of asymmetry and birefringence different from zero is plotted in Fig.~\ref{asymmetric}. Clearly, the effect of these terms is to lift the  degeneracy among the two central modes, thus breaking the symmetry of the spectra and inducing polaritons to polarise along a preferred direction. This is consistent with what it is observed in the experiments, although in the experiments the shape of the high-energy mode is generally less deformed by the asymmetry and birefringence than the low-energy modes. A possible explanation for this is that our theoretical model is based on the approximation of quadratic dispersion while in the polariton system the dispersion is strongly dependent on $k$. Since the modes are strongly confined, high $k$ vectors are likely to play an important role. In addition exciton-exciton interactions and pump-decay mechanisms, both of which are not included in our model, may also lead to experimental/theory differences in the details of the patterns.

\newpage

\section{Spin textures generated from higher order LG modes}

\begin{figure}
\center
\includegraphics [scale=1.2] {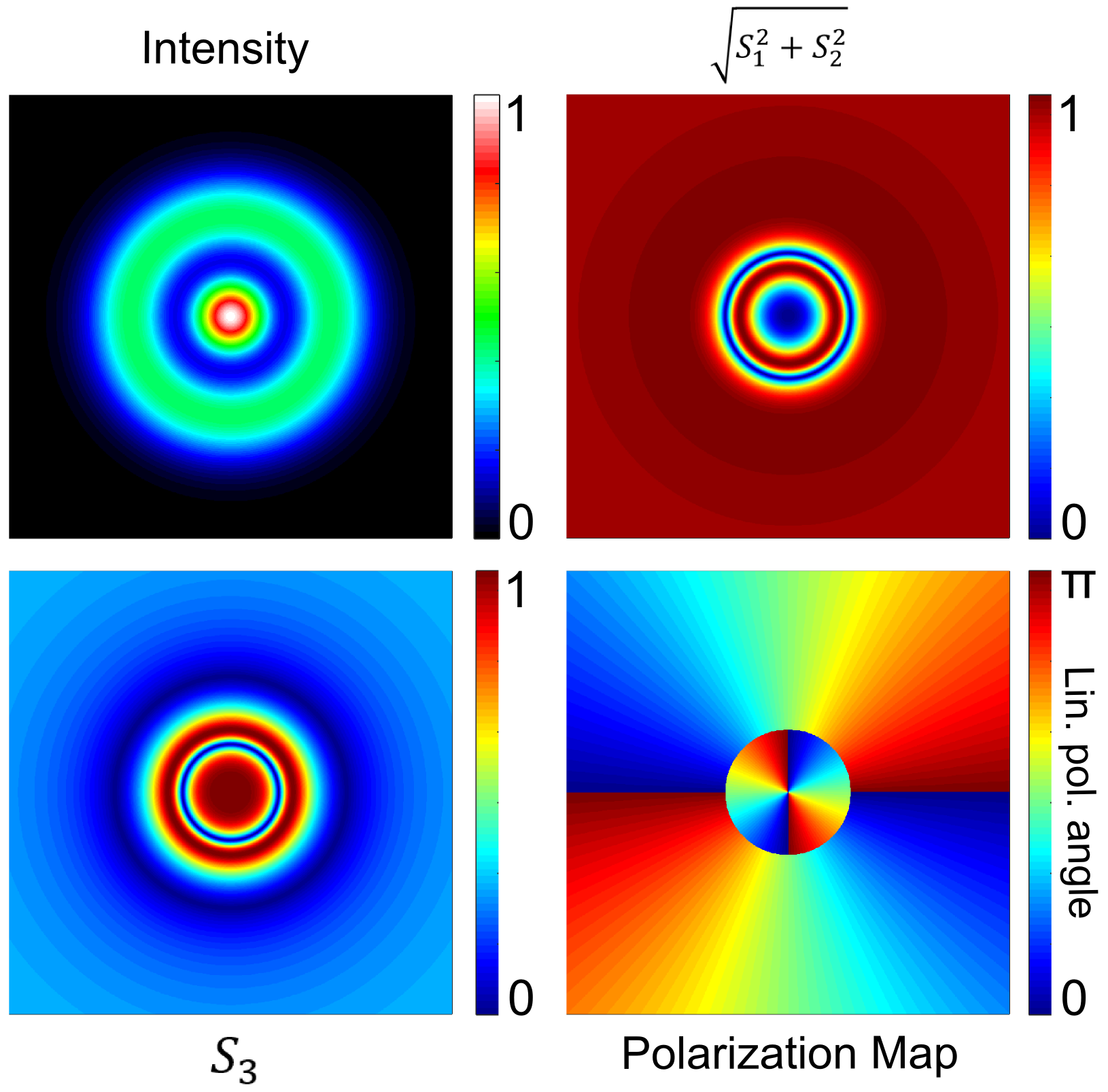}
\caption {\label{highorder} Simulated characters of the quasi-spin vortex A of Fig. 3 of the main text, including amplitude, linear polarization degree $\sqrt{S_{1}^{2}+S_{2}^{2}}$, circular polarization degree $S_{3}$ and linear polarization angle $arctan(S_{2}/S_{1})/2$. These characters illustrate the corresponding polarization pattern presented in the main text.}   
\end{figure}

We derive the polarization patterns by superposing  $\sigma^{+}  LG_{10}$ and $\sigma^{-} LG_{02}$. The result for the type A quasi-spin vortex in Fig.3 of the main text is shown in Fig.~\ref{highorder} as an example. The polarization pattern shows a radial spin vortex in the inner core and an azimuthal spin vortex in the outer ring, in good qualitative agreement with the experimental data shown in Fig. 3 (b) of the main text.

%\newpage

%\bibliography{Supplementary}

\end{document}